\documentclass[12pt,preprint]{aastex}

\newcommand{\sm}{M_\odot}
\newcommand{\sr}{R_\odot}
\newcommand{\RNum}[1]{\uppercase\expandafter{\romannumeral #1\relax}}

\def\farcsec{\hbox{$\ \!\!^{\prime\prime}$}}
\shorttitle{iPTF13bvn}
\shortauthors{Cao et al.}

\def\gtorder{\mathrel{\raise.3ex\hbox{$>$}\mkern-14mu
             \lower0.6ex\hbox{$\sim$}}}
\def\ltorder{\mathrel{\raise.3ex\hbox{$<$}\mkern-14mu
             \lower0.6ex\hbox{$\sim$}}}

\begin{document}

\title{Discovery, Progenitor \& Early Evolution of a Stripped Envelope Supernova \lowercase{i}PTF\lowercase{13bvn}}
\author{Yi~Cao\altaffilmark{1}, Mansi~M.~Kasliwal\altaffilmark{2}, Iair Arcavi\altaffilmark{3}, Assaf Horesh\altaffilmark{1}, Paul Hancock\altaffilmark{4,5},
              Stefano Valenti\altaffilmark{6,7}, S.~Bradley~Cenko\altaffilmark{8}, S.~R.~Kulkarni\altaffilmark{1},
              Avishay Gal-Yam\altaffilmark{3}, Evgeny~Gorbikov\altaffilmark{3},
              Eran~O.~Ofek\altaffilmark{3}, David~Sand\altaffilmark{9}, Ofer Yaron\altaffilmark{3}, 
              Melissa~Graham\altaffilmark{6,7}, Jeffrey~M.~Silverman\altaffilmark{10}, J.~Craig~Wheeler\altaffilmark{10},
              G.~H.~Marion\altaffilmark{10}, Emma~S.~Walker\altaffilmark{11}, Paolo~Mazzali\altaffilmark{12,19,20},
              D.~Andrew~Howell\altaffilmark{6,7}, K.~L.~Li\altaffilmark{18}, A.~K.~H.~Kong\altaffilmark{18}, 
             Joshua~S.~Bloom\altaffilmark{13}, Peter~E.~Nugent\altaffilmark{13,14}, 
              Jason~Surace\altaffilmark{15}, Frank~Masci\altaffilmark{16},
              John~Carpenter\altaffilmark{1}, Nathalie~Degenaar\altaffilmark{17}
                \&
                Christopher~R.~Gelino\altaffilmark{16}}
\altaffiltext{1}{Astronomy Department, California Institute of Technology, 1200 E. California Blvd., Pasadena, CA 91125, USA; ycao@astro.caltech.edu}
\altaffiltext{2}{The Observatories, Carnegie Institution for Science, 813, Santa Barbara Street, Pasadena, CA 91101, USA}
\altaffiltext{3}{Dept. of Particle Physics and Astrophysics, Weizmann Institute of Science, Rehovot, 76100, Israel} 
\altaffiltext{4}{Sydney Institute for Astronomy (SIfA), School of Physics, The University of Sydney, NSW 2006, Australia}
\altaffiltext{5}{ARC Centre of Excellence for All-sky Astrophysics (CAASTRO), The University of Sydney, NSW 2006, Australia}
\altaffiltext{6}{Las Cumbres Observatory Global Telescope Network, Goleta, CA 93117, USA}
\altaffiltext{7}{Department of Physics, University of California, Santa Barbara, CA 93106, USA}
\altaffiltext{8}{Astrophysics Science Division, NASA Goddard Space Flight Center, Mail Code 661, Greenbelt, MD, 20771, USA}
\altaffiltext{9}{Department of Physics, Texas Tech University, Lubbock, TX 79409, USA}
\altaffiltext{10}{Department of Astronomy, University of Texas at Austin, Austin, TX 78712, USA}
\altaffiltext{11}{Department of Physics, Yale University, New Haven, CT 06511-8499, USA}
\altaffiltext{12}{INAF-Padova Astronomical Observatory, Vicolo dell'Osservatorio 5 35122 Padova Italy}
\altaffiltext{13}{Department of Astronomy, University of California Berkeley, B-20 Hearst Field Annex \# 3411, Berkeley, CA, 94720-3411, USA}
\altaffiltext{14}{Computational Cosmology Center, Computational Research Division, Lawrence Berkeley National Laboratory, 1 Cyclotron Road MS 50B-4206, Berkeley, CA, 94720, USA}
\altaffiltext{15}{Spitzer Science Center, MS 220-6, California Institute of Technology, Jet Propulsion Laboratory, Pasadena, CA 91125, USA}
\altaffiltext{16}{Infrared Processing and Analysis Center, California Institute of Technology, Pasadena, CA 91125, USA}
\altaffiltext{17}{Department of Astronomy, University of Michigan, 500 Church Street, Ann Arbor, MI 48109, USA}
\altaffiltext{18}{Institute of Astronomy and Department of Physics, National Tsing Hua University, Hsinchu 30013, Taiwan}
\altaffiltext{19}{Astrophysics Research Institute, Liverpool John Moores University, Liverpool, UK}
\altaffiltext{20}{Max-Planck Institute for Astrophysics, Garching, Germany}

\begin{abstract}
The intermediate Palomar Transient Factory reports our discovery of a young supernova, iPTF13bvn, in the nearby galaxy, NGC\,5806 (22.5\,Mpc). 
Our spectral sequence in the optical and infrared suggests a Type \RNum{1}b classification. 
We identify a blue progenitor candidate in deep pre-explosion imaging within a 2$\sigma$ error circle of 80\,mas (8.7\,pc). 
The candidate has a M$_B$ luminosity of $-$5.52$\,\pm\,$0.39\,mag and a $B-I$ color of 0.25$\,\pm\,$0.25\,mag.  
If confirmed by future observations, this would be the first direct detection for a progenitor of a Type \RNum{1}b. 
Fitting a power law to the early light curve, we find an extrapolated explosion date around 0.6\,days before our first detection. 
We see no evidence of shock cooling. The pre-explosion detection limits constrain the radius of the progenitor to be smaller 
than a few solar radii.
iPTF13bvn is also detected in cm and mm-wavelengths. Fitting a synchrotron self-absorption model to our radio data, 
we find a mass loading parameter of 1.3$\times$10$^{12}$ g cm$^{-1}$. Assuming a wind velocity of $10^{3}\,$km\,s$^{-1}$, 
we derive a progenitor mass loss rate of $3\times10^{-5} \sm$ yr$^{-1}$.
Our observations, taken as a whole, are consistent with a Wolf Rayet progenitor of the supernova iPTF13bvn. 
\end{abstract}

\keywords{supernovae: individual (iPTF13bvn) --- surveys --- stars: Wolf-Rayet --- shock waves --- instrumentation: adaptive optics}

\section{Introduction}
Supernovae of Type \RNum{1}b/c constitute about one third of the
death rate of massive stars \citep{LCL2011}. Their spectra lack hydrogen \citep{Filippenko1997}, 
suggesting progenitors stripped of their hydrogen envelopes either due to mass transfer in a binary system or via
copious stellar winds. On theoretical grounds,
the anticipated progenitors are Wolf Rayet (WR) stars or massive Helium stars 
\citep[e.g.,][]{DHL2012, YGV2012}.

The most direct way to test the above theoretical picture is
direct detection of the progenitor. This is possible for supernovae occurring 
in nearby galaxies which have deep pre-explosion images (which means, in practice, \textit{HST}). 
\citet{EFS2013} comprehensively summarize searches for twelve \RNum{1}b/c progenitors, which are all non-detections.  
The deepest upper limit to date is $M_B=-4.4$\,mag for SN\,2002ap, a Type \RNum{1}c supernova in the very
nearby galaxy Messier~74 \citep{CSE2007}.

An indirect way to infer the progenitor properties is the early light curve of a supernova, 
which is sensitive to the progenitor size, explosion energy and the composition of the outer
layers \citep[e.g.,][]{NS2010, RW2011, BBN2012, PN2013}, and in some propitious cases can even 
diagnose the presence of a binary companion \citep{Kasen2010}.
No detection of shock cooling constrains the radius of PTF10vgv (Type \RNum{1}c) to $<6\sr$ \citep{COG2012,PN2013}.
For SN2008D (Type \RNum{1}b), the shock cooling phase lasted about five days and the inferred
radius is $<14\,\sr$ \citep{SBP2008}.

Radio measurements serve as another indirect probe of the progenitor system by 
characterizing the mass-loss history. Radio emission is produced by relativistic electrons accelerated 
in the supernova shock as they gyrate in the amplified magnetic field when 
the shock expands freely. The radio spectral evolution implies a mass
loading parameter $A=\dot{\rm M}/(4\pi v_w)$ where $\dot{\rm M}$ is a constant mass-loss rate 
and $v_w$ is the wind velocity from the progenitor \citep{Chevalier1998}. 
 Type \,\RNum{1}b/c supernovae typically have $A$ of the order of 10$^{12}$ g cm$^{-1}$ \citep{CF2006}.

In this letter, we report the discovery of iPTF13bvn, a young Type 
\RNum{1}b supernova in the nearby galaxy NGC\,5806. Archival \textit{HST} images of NGC\,5806 allow a direct progenitor search.
We also present early photometric, spectroscopic and wide-band radio observations and discuss implications on the progenitor.  

\section{Discovery}
\label{sec:discovery}
\begin{figure}
\centering
\includegraphics[width=0.8\textwidth]{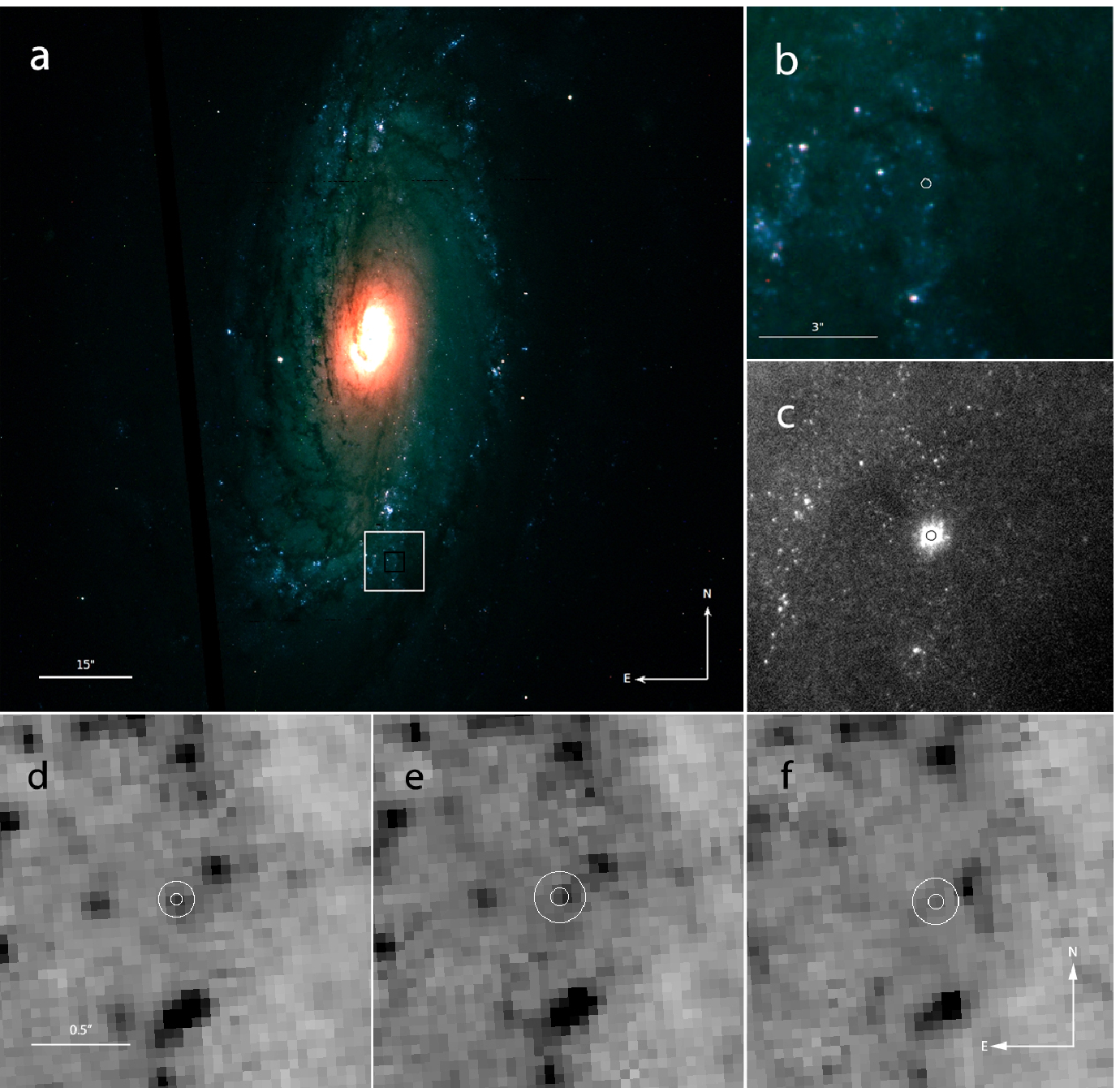}
\caption{Supernova iPTF13bvn and its host galaxy, NGC5806. The supernova is located
at $\alpha = 15^{\mathrm{h}} 00^{\mathrm{m}} 00.152^{\mathrm{s}}$,
$\delta = +01^{\circ} 52\arcmin 53.\farcsec17$ (J2000).
Panel (a) shows the \textit{HST} image of NGC\,5806, taken on 2005-03-10 UT (\textit{HST} proposal 10187, PI Smartt).  Panel (b)
is a zoom-in of the \textit{HST} image near the site of iPTF13bvn. Panel (c) is the composite image of the supernova taken with OSIRIS
and LGS-AO system. This is made by stacking sixty 15\,s exposures. We align this image with the \textit{HST}/ACS
images with 15, 23, 25 registration stars in the F435W, F555W and F814W filters and acquire registration uncertainties of 0.6, 0.9 and 0.8 
\textit{HST} pixel, respectively. Panel (d), (e) and (f) show the \textit{HST} images at the position of the 
supernova in F435W, F555W and F814W filters. The position of 
iPTF13bvn is marked with 1-$\sigma$ and 3-$\sigma$ error circles. 
The progenitor candidate is 
$26.50\pm0.15$\,mag in F435W ($S/N=11.2$), $26.40\pm0.15$\,mag in F555W ($S/N=9.9$) and $26.10\pm0.20$\,mag in F814W ($S/N=8.2$). 
\label{fig:discovery}}
\end{figure}

On 2013 Jun 16.238 UT{\footnote{All times are in UT.}},  the automated real-time discovery
and classification pipeline of the intermediate Palomar Transient Factory
\citep[iPTF;][Nugent et al. in prep]{LKD2009} identified a
new transient source with $r$=18.6\,mag{\footnote{BVRI mags are in Vega; 
all other mags are in AB system.}} 
in the vicinity of NGC\,5806 (see Figure \ref{fig:discovery}, \citealt{ATel5137}).
No source was detected at the same location to $<21.7$\,mag (3$\sigma$) on Jun 15.240 (Figure~\ref{fig:lc}).  
Also, there was no evidence for pre-outburst activity in 492 PTF images taken since 2009 Jun 29 to a similar depth.
Our duty astronomer saved this source as iPTF13bvn and initiated rapid, 
multi-wavelength follow-up.

\section{Progenitor Identification}
\label{sec:prog}
On Jun 20.276, we observed iPTF13bvn in $H$-band with
OSIRIS \citep{LBK2006} and the Laser Guide Star Adaptive Optics (LGS-AO) system \citep{WCJ2006} mounted on the 10\,m Keck 
\RNum{1} telescope.
Registering the AO image to the archival \textit{HST}/ACS image, we get a 1$\sigma$ uncertainty of $\lessapprox$40\,mas.
We find one source coincident with the supernova within the 2$\sigma$ (equivalently 8.7\,pc projection distance) error circle 
(Figure~\ref{fig:discovery}; \citealt{ATel5152}). 

We perform PSF photometry on the {\it HST}/ACS images with DOLPHOT \citep{Dolphin2000}. The photometry of the progenitor candidate is 
$26.50\pm0.15$\,mag in F435W, $26.40\pm0.15$\,mag in F555W and $26.10\pm0.20$\,mag in F814W.
To correct for extinction, we obtain a high resolution spectrum (\S\,\ref{sec:obs}) and measure the equivalent widths of Na \RNum{1} D lines. 
We find local extinction of $E(B-V)=0.0437$ and foreground extinction of 0.0278 \citep{PPB2012}. 
Assuming $R_V=3.1$ \citep{SF2011} and adopting a distance modulus of 31.76$\pm0.36$ (22.5\,Mpc;  \citealt{TRS2009}), 
we find ${\rm M_B=-5.52\pm0.39}$\,mag, ${\rm M_V=-5.55\pm0.39}$\,mag 
and ${\rm M_I=-5.77\pm0.41}$\,mag. Thus the progenitor, if single, is no brighter than these values. The luminosity and colors are consistent with 
the compilation of WN and WC Wolf Rayet stars in \citet{EFS2013}.

However, WR stars are often in binaries \citep{TMD1999}. Thus, the progenitor is possibly in a binary system
and the light is dominated by the companion. We further note that color alone cannot be used as a discriminant. 
The B, V and I filters are in the Rayleigh-Jeans tail of hot stars, including O-stars, WR stars and blue supergiants. 
Moreover, given the $0\farcsec.05$ pixel size of {\it HST}/ACS, or equivalently 5.45\,pc at the distance of NGC\,5806, the candidate can also be an
unresolved young star cluster whose color is dominated by OB stars. Finally, we caution that
the progenitor candidate may even be unrelated to the supernova.
 
The litmus test of whether this candidate is the progenitor or part of the progenitor system
can only be undertaken by \textit{HST} imaging after the supernova fades.

\section{Early Photometric and Spectroscopic Evolution}
\label{sec:evol}

\subsection{Observations and Reduction}
\label{sec:obs}
As part of regular iPTF operations, the field of iPTF13bvn was imaged twice a night every night 
by the Palomar 48-inch Oschin telescope (P48) with a Mould $R$-band filter \citep{oll+12} during the Spring quarter. 
Upon discovery of iPTF13bvn, the robotic Palomar 60-inch telescope \citep[P60;][]{CFM2006}  
was triggered for follow-up in $g^{\prime}$$r^{\prime}$$i^{\prime}$$z^{\prime}$-bands.
We obtained photometry with the Las Cumbres Observatory Global Telescope (LCOGT; \citealt{BBB2013}) 
network in 
$UBVRIg^{\prime}r^{\prime}i^{\prime}z^{\prime}$-bands using the 1-m telescopes from Cerro
Tololo (Chile), McDonald Observatory (USA) and Sutherland (South
Africa), along with the 2m Faulkes Telescope South (Siding
Springs). As part of our ongoing iPTF-\textit{Swift} program{\footnote{PID 9120112, PI Kasliwal}},
we triggered target-of-opportunity observations beginning on 2013 Jun 17 \citep{ATel5146}. 
In the P48, P60 and \textit{Swift} images, the host background is subtracted by using pre-explosion reference images, while
in the LCOGT images a low-order polynomial fit is used to remove the background. PSF photometry is then
performed in all the images. Photometry in $g^{\prime}r^{\prime}i^{\prime}z^{\prime}$-bands is calibrated to SDSS stars and that in
$UBVRI$-bands is calibrated with Landolt standard stars. 
The multi-color light curve of iPTF13bvn is illustrated in Figure~\ref{fig:lc}. 

\begin{figure}
\begin{center}
\includegraphics[width=0.8\textwidth]{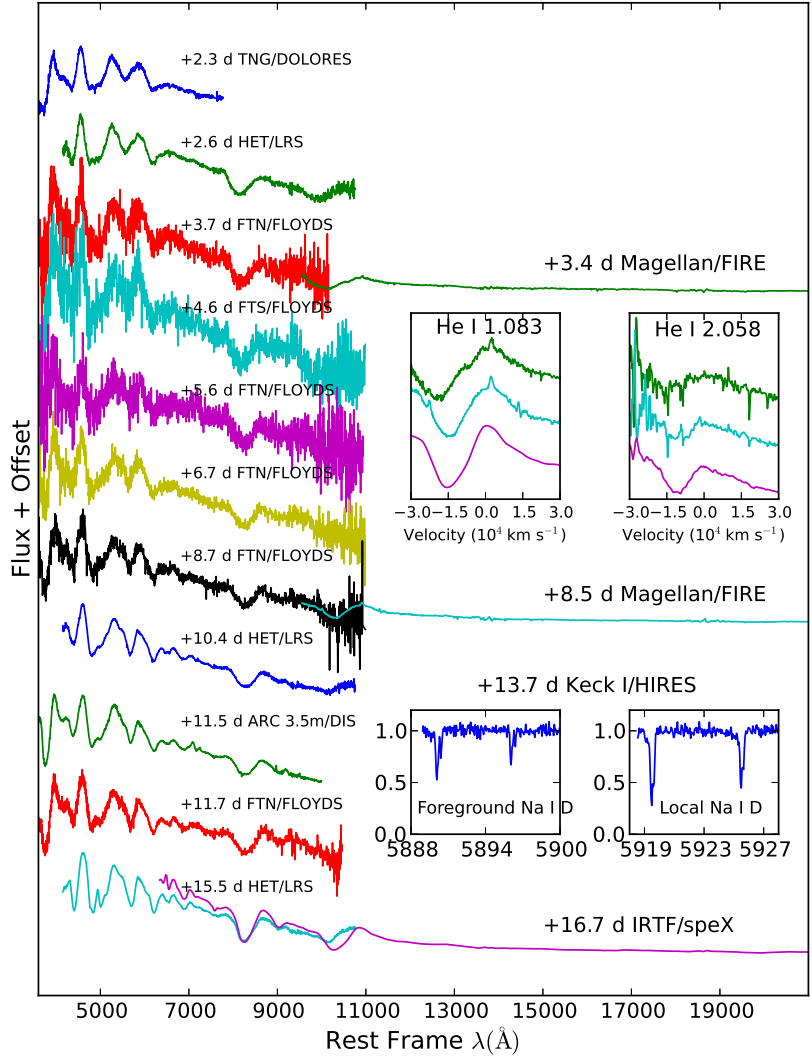}
\caption{The spectral sequence of iPTF13bvn. Each spectrum is labelled with the observation date, the telescope and the instrument. 
The upper two insets zoom-in to the infrared He \RNum{1} 10830 and He \RNum{1} 20581 lines. 
The lower two insets zoom-in to the Na \RNum{1} D doublet at high resolution. 
\label{fig:spec_seq}}
\end{center}
\end{figure}

Low-resolution spectroscopic follow-up of iPTF13bvn was undertaken with the DOLORES low-resolution spectrograph on 
Telescopio Nazionale Galileo (TNG), the Marcario Low-Resolution Spectrograph (LRS; \citealt{Hill98}) on the Hobby-Eberly Telescope (HET),
the low-resolution, cross-dispersed spectrograph FLOYDS (Sand et al. in prep) on the robotic Faulkes Telescope (FT), 
the Dual Imaging Spectrograph (DIS) on the ARC 3.5-m telescope, the Folded-Port InfraRed Echellete \citep[FIRE,][]{SBS2013} 
on the 6.5-m Magellan telescope, and SpeX \citep{RTO2003} on the NASA Infrared Telescope Facility (IRTF). 
%The flux calibrators are Feige\,34 for the DOLORES and FLOYDS spectra, BD\,262606 for the LRS spectra
%and BD\,332642 for the DIS spectrum. The A0V standard stars, HD\,153068 and HD\,125062, 
%were used for flux calibration and removal of telluric absorption features 
%in the FIRE spectra. 
We also obtained a high-resolution spectrum with High Resolution Echelle Spectrometer \citep[HIRES;][]{VAB1994} on the 10\,m 
Keck I telescope. The spectroscopic series of
iPTF13bvn is displayed in Figure~\ref{fig:spec_seq}. 

All photometry tables and spectroscopy data will be made public via 
WISeREP{\footnote{http://www.weizmann.ac.il/astrophysics/wiserep/}} \citep{YG2012}.

\subsection{Analysis: Photometry}
\begin{figure}
\begin{center}
\includegraphics[width=0.8\textwidth]{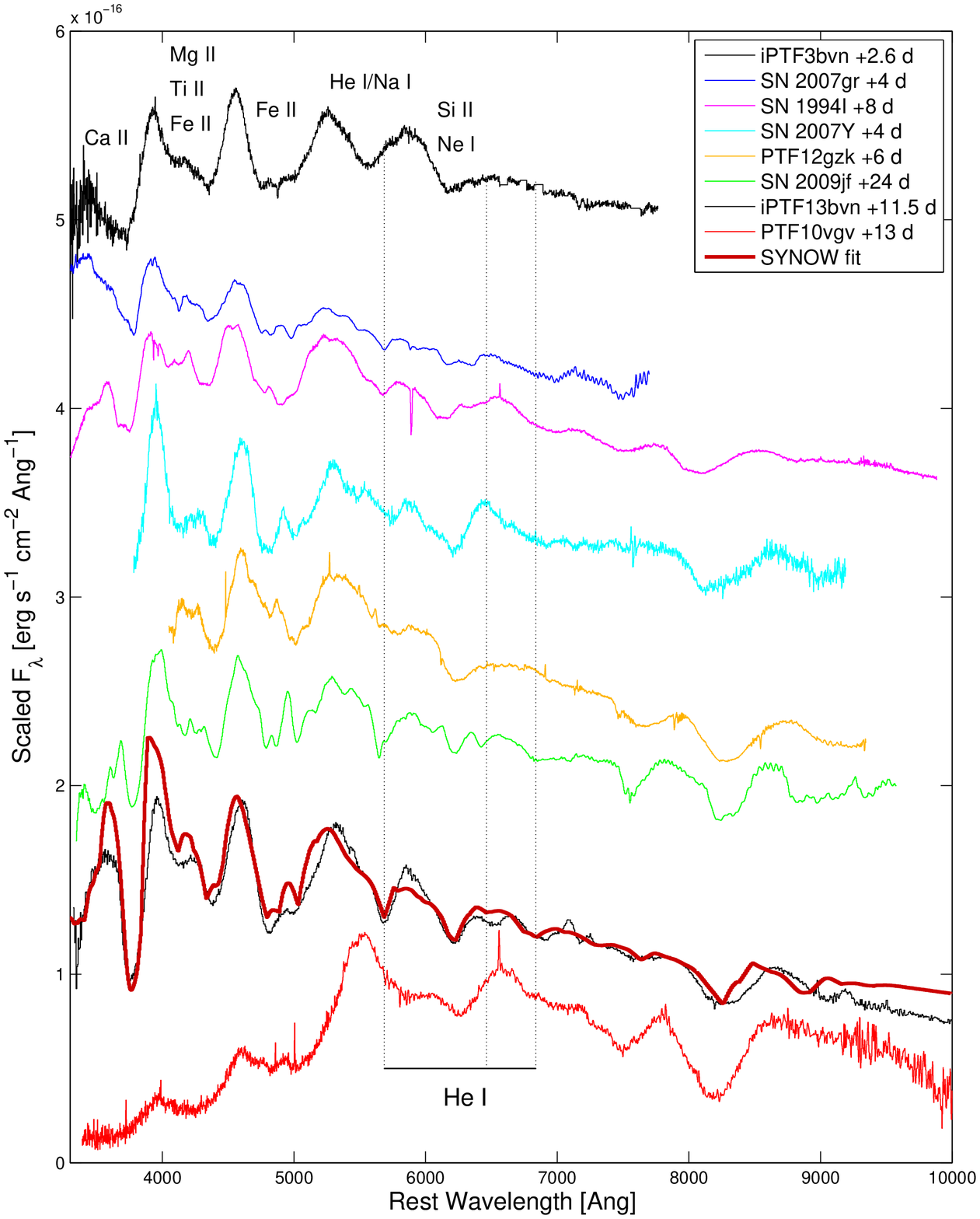}
\caption{The spectra of iPTF13bvn (black) compared to other SNe\,\RNum{1}b/c. All ages are reported in days since explosion. An early spectrum of iPTF13bvn shows strong similarity to early spectra of 
Type \RNum{1}b/c SNe: SN2007gr, Type \RNum{1}c \citep{VET2008}; SN1994I, Type Ic \citep{FBM1995}; SN2007Y, Type \RNum{1}b \citep{SMP2009}. 
The early spectrum of the energetic Type Ic PTF12gzk \citep{BGF2012} resembles that of iPTF13bvn,
if a correction is made artificially for its remarkable blueshift. 
Later spectra of iPTF13bvn resemble the Type Ib SN\,2009jf \citep{VFB2011}. Despite the light curve similarity, the high velocity Type Ic PTF10vgv bears no spectral resemblance to iPTF13bvn. 
 \label{fig:spec_comp}}
\end{center}
\end{figure}

In order to better constrain the explosion date, we fit a power-law model 
$f(t)\propto(t-t_0)^{\beta}$ to the P48 data 
of iPTF13bvn taken in the first five nights after explosion and constrain the parameters with the upper limits preceding
the discovery. The best fit results in an explosion date 
$t_0=$ Jun 15.67 and a power law index $\beta=1.01$ (Figure \ref{fig:spec_comp}). The $95\%$ confidence levels for
$t_0$ and $\beta$ are $[{\rm Jun\,15.50, Jun\,15.76}]$ and $[0.923, 1.09]$, respectively. 
Note that the color evolution is minimal, suggesting that r-band is a reasonable proxy for the bolometric light curve. 
iPTF13bvn peaked at $r\,=\,-$16.6\,mag at about $t_0\,+\,$18\,days.

\begin{figure}
\includegraphics[width=\textwidth]{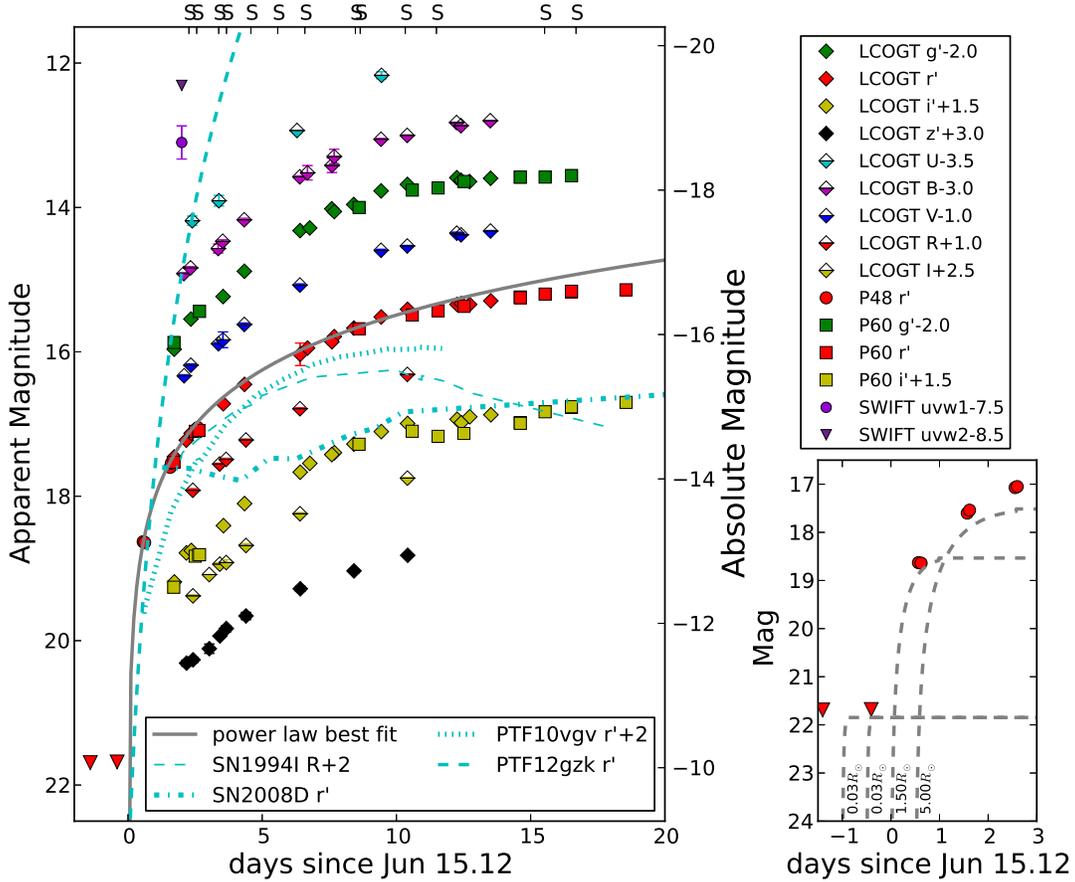}
\caption{The left panel shows the multi-band light curve of iPTF13bvn (color represents filter, symbol shape represents telescope). 
The upper limits are denoted in triangles. 
The origin $t=0$ is set to the explosion date derived from the best power-law
fit (solid gray curve). For comparison, $r$-band light curves of 
SN\,1994I, SN\,2008D, PTF10vgv and PTF12gzk are also shown in cyan with different line styles. 
On the top axis, the epochs of spectroscopic follow-up are indicated by ``S".  
In the right panel, the P48 $r$-band light curves are plotted against predicted light curves of shock cooling (gray dash curves) from \citet{PN2013}
for explosion starting at $t=-1\ ,\ -0.5,\ 0.0,\ 0.5$\,days with a ``mean'' velocity of $8,000\,$km\,s$^{-1}$. The radius of the progenitor in each of the models 
is besides its corresponding light curve. 
 \label{fig:lc}}
\end{figure}

A direct comparison of the iPTF13bvn $r$-band light curve with 
SN\,1994I \citep{RvH1996}, SN\,2008D \citep{SBP2008}, PTF10vgv \citep{COG2012} 
and PTF12gzk \citep{BGF2012} is also shown in Figure \ref{fig:lc}. The Type \RNum{1}c 
SN\,1994I is much more luminous than iPTF13bvn
while its rise rate is less than that of iPTF13bvn. The Type \RNum{1}b SN\,2008D has a five day cooling 
phase after its shock breakout before it slowly rises again,  
while we do not detect any shock cooling signature from iPTF13bvn. The energetic Type Ic 
supernova PTF12gzk rises much faster than iPTF13bvn. Surprisingly, the closest match to the 
rise rate of iPTF13bvn through 8\,days is the high velocity Type Ic PTF10vgv (albeit about two magnitudes fainter).

%R-band light curve follows bolometric
Following \citet{PN2013}, we derive a constraint on the progenitor radius based on our detection limits on
shock cooling. The inferred radius is sensitive to the ``mean'' velocity of the ejecta 
$v=\sqrt{2E/M_{\rm ej}}$ where $E$ is the explosion energy and $M_{\rm ej}$ is the mass
of the ejecta. Based on the measured photospheric velocities and line velocities (\S\,\ref{sec:spec}), we conservatively adopt the lowest velocity of 
8000\,km\,s$^{-1}$.  If the velocity was higher, the radius of the progenitor would be constrained to be even smaller \citep{PN2013}.
The model also depends weakly on $E$ and thus we assume $E=10^{51}$\,erg. 
In the right bottom panel of Figure~\ref{fig:lc}, the predicted $r$-band light curves (gray lines) of
shock cooling with various explosion times are also plotted. We find: if the explosion had happened
earlier than t$_0$, the progenitor radius would have to be unreasonably small 
($\approx\,0.03\sr$) because the luminosity of the shock breakout is tightly constrained by non-detection on Jun 14. 
If the explosion happened at t$_0$, the progenitor radius $<\,1.5\sr$. 
If the explosion happened at t$_0$ + 0.5\,day, the progenitor radius $<\,5\sr$.
We conclude that the progenitor radius was no larger than a few solar radii, suggesting a stripped core such as WR stars. 

\subsection{Analysis: Spectroscopy}
\label{sec:spec}

The strongest features in the early spectra are the Ca \RNum{2} H$+$K and Ca \RNum{2} near-IR triplet absorption. 
The blends of Fe \RNum{2} lines at $\approx 5000$\AA\ and Fe \RNum{2}, Mg \RNum{2}
and Ti \RNum{2} at $\approx 4400$\AA\ are also prominent. The local minimum at about $5500$\AA\ may be either He \RNum{1}
or Na \RNum{1} and that at about $6200$\AA\ may be interpreted as either Si \RNum{2} or Ne \RNum{1}. 
At +2.6\,days after explosion, a SYNOW fit gives a photospheric velocity of 10000\,km\,s$^{-1}$ and a Ca II line velocity of
14000\,km\,s$^{-1}$. By +11.5\,days, these velocities evolve to 8000 and 10000\,km\,s$^{-1}$ respectively. 

Helium lines are not expected to be prominent in early spectra of Type Ib supernovae \citep{HMT2012}. 
We begin to see weak wiggles at the location of He I (5876\AA\, 6678\AA\ and 7065\AA)
in optical spectra after +10\,days. However, SYNOW cannot fit a single expansion velocity to all the three lines simultaneously.
In the IR spectra, we see a prominent 10200\,\AA\ absorption feature, which may be explained by He \RNum{1} 10830 line blended with metal lines. 
The 20581\,\AA\  doublet is often used to confirm the presence of He \citep{TPM2006,MVK2013}.
By +16.7\,days, we unambiguously detect this feature (see the inset of Figure~\ref{fig:spec_seq}). 
Therefore, we classify iPTF13bvn as a Type \RNum{1}b supernova \citep{ATel5142, ATel5151}.

\section{Radio Follow-up}
\label{sec:radio}
\begin{figure}
\centering
\includegraphics[width=0.9\textwidth]{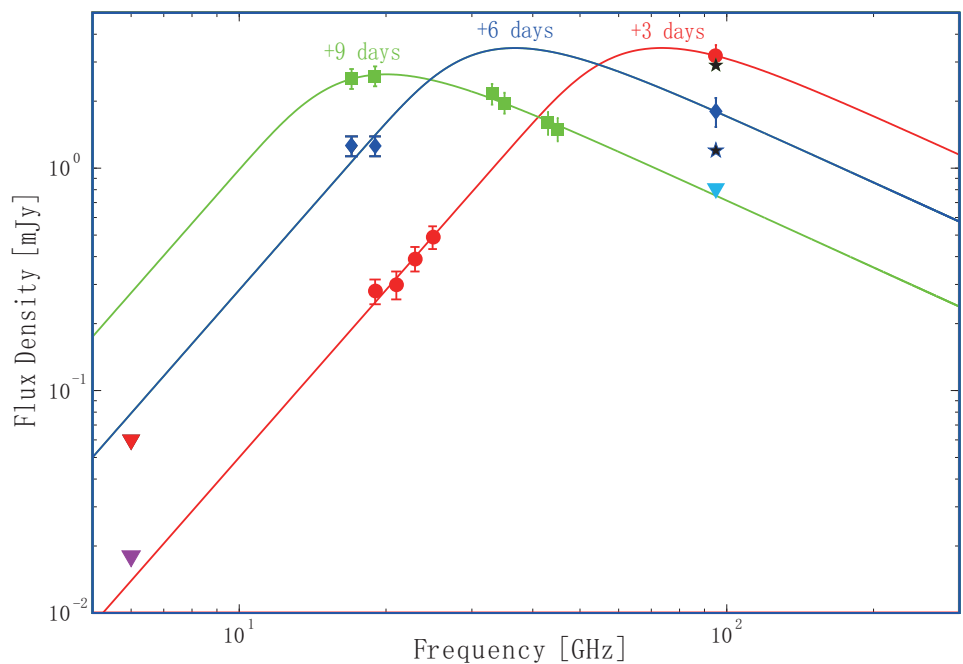}
\caption{ Flux density as a function of frequency on +3\,d (red circles; CARMA + JVLA), 
+6\,d (blue diamonds; CARMA + ATCA), and +9\,d (green squares; ATCA). 
The red triangle is our JVLA detection limit ($3\sigma$) in C band (6\,GHz) at +3\,d. 
The purple triangle is the JVLA detection limit ($3\sigma$) on +4\,d obtained by \citet{ATel5154}.
The solid lines show our fits using synchrotron self-absorption models.
Additional CARMA measurements on +4\,d, and +8\,d  (black stars) and an upper limit on +10\,d (cyan triangle) are shown.
\label{fig:radio}}
\end{figure}

Promptly after discovery, we initiated radio follow-up
observations at the Combined Array for Research in Millimeter-wave
Astronomy (CARMA)\footnote{CARMA program c1127, PI Horesh}
and requested Director's Discretionary Time at the Jansky Very Large
Array (JVLA)\footnote{JVLA program 13A-531, PI Horesh}. Our JVLA and CARMA data showed that the radio
counterpart had a synchrotron self-absorption (SSA) spectrum. On being denied our request
for additional JVLA observations, our colleagues were able to initiate Australia Telescope Compact
Array (ATCA)\footnote{ATCA program C2707, PI Hancock } observations. 
The JVLA data were reduced using AIPS with
J$1505+0326$ as a phase calibrator and 3C286 as a flux calibrator. The CARMA data were 
reduced using MIRIAD \citep{STW1995} with J$1549+026$ as a phase calibrator and MWC349 
as a flux calibrator. The ATCA data were reduced using MIRIAD with 
PKS\,B1921-293 as a bandpass calibrator, PKS\,B1508-055 
as a phase calibrator and PKS\,B1934-638 as a flux calibrator. 
All the radio data is presented in Figure~\ref{fig:radio}.

We fit the multi-frequency multi-epoch data to the SSA model following \citet{Chevalier1998}. 
We assume that the electron energy
distribution can be described by a power law $N(E)\propto E^{-3}$. The fitted models are also presented in
Figure~\ref{fig:radio}. 

We note that the SSA model alone is 
consistent with all our data including the null-detection in C band on Jun 18. However, on Jun 19.1, \citet{ATel5154} observed
iPTF13bvn with the JVLA in C band and announced  a null-detection with an
rms of $6\,\mu$Jy. Based on our SSA model, we estimate that Kamble \& Soderberg 
should have detected iPTF13bvn at a level of $\sim
5\sigma$. Their null detection therefore may require
additional free-free absorption at low frequencies. 
Based on inverse Compton scattering, we estimate the X-ray emission to be $10^{-15}\,{\rm erg\,cm^{-2}\,s^{-1}}$. 
This is consistent with our \textit{Swift}/XRT upper limit of $1.0\times10^{-13}\,{\rm erg\ cm^{-2}\ s^{-1}}$ on Jun 17 
(assuming a power-law spectrum with a photon index of 2.0). We note that \citet{ATel5210} reported an X-ray detection by
co-adding the XRT data from Jun 17 to Jul 13. We reprocessed the XRT data from the first ten days after the supernova explosion and resulted in
a 3-$\sigma$ upper limit of $4\times10^{-14}\,{\rm erg\,cm^{-2}\,s^{-1}}$. This is also consistent with our expectation from inverse Compton
scattering. 

We find that the radio emission peaks at
$\approx\,73.5,\ 37,\ 20$\,GHz with fluxes of $\approx 3.5,\ 3.5,\ 2.7$\,mJy on Jun 18, 
21 and 24, respectively. Following \citet{Chevalier1998}, we derive
a shockwave radius $R_{s}\approx\,0.7,\ 1.4,\ 2.3 \times 10^{15}\,{\rm cm}$ and a magnetic
field strength $B\approx\,8.2,\ 4.1,\ 2.3\,{\rm G}$. Using the explosion date determined in the optical data, 
we find a shock wave velocity of $2.7\times 10^4\,{\rm km\,s^{-1}}$, which is typical for SNe\,\RNum{1}b/c \citep{CF2006}. 

We derive a mass-loading parameter $A\approx 1.3 \times
10^{12}$\,g\,cm$^{-1}$  \citep[see][for equations]{HKF2012}.
Assuming a typical wind velocity of $v_{w}\approx\,$1000 km s$^{-1}$ from a WR star \citep{CGv2004}, 
the progenitor mass loss rate $\dot{\rm M}=4\pi Av_w \approx\,3\times10^{-5}\,\sm\,{\rm yr^{-1}}$.
This is consistent with the WR mass loss rate in \citet{CGv2004}.

\section{Conclusion}
\label{sec:con}
iPTF13bvn is a Type \RNum{1}b supernova that exploded on Jun 15.67
and rose to maximum luminosity of M$_R=-$16.6 in about 18\,days. The luminosity of the supernova in the first ten days is approximately
proportional to $t^{-1}$. 
We identify a single progenitor candidate within a 2$\sigma$ error radius of 8.7\,pc in pre-explosion \textit{HST} imaging. 
The candidate has a M$_B$ luminosity of $-$5.52$\,\pm\,$0.39\,mag and a $B-I$ color of 0.25$\,\pm\,$0.25\,mag.
Future \textit{HST} imaging, after the supernova fades away, will determine whether this is a single star, a binary or a star cluster. 
The non-detection of shock cooling in our light curve constrains the progenitor radius to smaller than a few solar radii. 
Our radio detections suggest a shock wave with velocity of $2.7\times10^4$\,km s$^{-1}$ and a progenitor
mass loss rate of $3\times10^{-5}\sm\,{\rm yr^{-1}}$. 

We conclude that the pre-explosion photometry of the detected candidate, the radius constraint based on absence of shock cooling, 
and the mass loss rate derived from radio are all consistent with a Wolf Rayet progenitor.

\section*{Acknowledgement}
We thank A. L. Piro for valuable discussions. We thank the following people for co-operating with our Target Of Opportunity or queue observations:
M. Roth (Magellan), A. Hartuynan (TNG), J. Johnson (Keck), J. Caldwell (HET).  We thank A. Howard and H. Isaacson for
HIRES data reduction. We thank R. Campbell, 
Hien Tran and S. Tendulkar for helping with
OSIRIS LGS-AO observation and data reduction.
We thank P. Vreeswijk for assisting with
\textit{HST} image registration. We thank J. Vinko, R. Foley, B. Kirshner, D. Perley , A. Corsi and K. Mooley as proposal co-Is.
We thank J. Swift, B. Montet, M. Bryan, R. Jensen-Clem, D. Polishook, S. Tinyanont for assisting with observations. 

M.~M.~K. acknowledges generous support from the Hubble Fellowship and Carnegie-Princeton Fellowship. 
JMS is supported by an NSF Astronomy and Astrophysics Postdoctoral
Fellowship under award AST-1302771. N.~D. acknowledges the Hubble Fellowship. 
Research by AGY and his group was supported by grants
from the ISF, BSF, GIF, Minerva, the EU/FP7 via an ERC grant and the Kimmel award.
The research of JCW is supported by NSF Grant AST 11-09801.
 
The National Energy Research Scientific Computing Center, supported by the Office of Science of the U.S. Department of Energy, provided staff, computational resources, and data storage for this project.
The Australia Telescope is funded by the Commonwealth of Australia for operation as a National Facility managed by CSIRO.
This research has been supported by the Australian Research Council through Super Science Fellowship grant
FS100100033. The Centre for All-sky Astrophysics is an Australian Research Council Centre of Excellence,
funded by grant CE110001020. The
National Radio Astronomy Observatory is a facility of
the National Science Foundation operated under cooperative agreement by Associated Universities, Inc. Ongoing CARMA development and operations are supported by the National Science Foundation under a cooperative agreement, and by the CARMA partner universities.

\begin{thebibliography}{51}
\expandafter\ifx\csname natexlab\endcsname\relax\def\natexlab#1{#1}\fi

\bibitem[{{Arcavi} {et~al.}(2013{\natexlab{a}}){Arcavi}, {Cenko}, {Gal-Yam}, \&
  {Ofek}}]{ATel5146}
{Arcavi}, I., {Cenko}, S.~B., {Gal-Yam}, A., \& {Ofek}, E. 2013{\natexlab{a}},
  The Astronomer's Telegram, 5146, 1

\bibitem[{{Arcavi} {et~al.}(2013{\natexlab{b}}){Arcavi}, {Ofek}, {Cao},
  {Gelino}, {Yaron}, {Vreeswijk}, {Gal-Yam}, {Cenko}, {Kong}, \&
  {Li}}]{ATel5152}
{Arcavi}, I., {et~al.} 2013{\natexlab{b}}, The Astronomer's Telegram, 5152, 1

\bibitem[{{Ben-Ami} {et~al.}(2012){Ben-Ami}, {Gal-Yam}, {Filippenko},
  {Mazzali}, {Modjaz}, {Yaron}, {Arcavi}, {Cenko}, {Horesh}, {Howell},
  {Graham}, {Horst}, {Im}, {Jeon}, {Kulkarni}, {Leonard}, {Perley}, {Pian},
  {Sand}, {Sullivan}, {Becker}, {Bersier}, {Bloom}, {Bottom}, {Brown}, {Clubb},
  {Dilday}, {Dixon}, {Fortinsky}, {Fox}, {Gonzalez}, {Harutyunyan}, {Kasliwal},
  {Li}, {Malkan}, {Manulis}, {Matheson}, {Moskovitz}, {Muirhead}, {Nugent},
  {Ofek}, {Quimby}, {Richards}, {Ross}, {Searcy}, {Silverman}, {Smith},
  {Vanderburg}, \& {Walker}}]{BGF2012}
{Ben-Ami}, S., {et~al.} 2012, \apjl, 760, L33

\bibitem[{{Bersten} {et~al.}(2012){Bersten}, {Benvenuto}, {Nomoto}, {Ergon},
  {Folatelli}, {Sollerman}, {Benetti}, {Botticella}, {Fraser}, {Kotak},
  {Maeda}, {Ochner}, \& {Tomasella}}]{BBN2012}
{Bersten}, M.~C., {et~al.} 2012, \apj, 757, 31

\bibitem[{{Brown} {et~al.}(2013){Brown}, {Baliber}, {Bianco}, {Bowman},
  {Burleson}, {Conway}, {Crellin}, {Depagne}, {De Vera}, {Dilday}, {Dragomir},
  {Dubberley}, {Eastman}, {Elphick}, {Falarski}, {Foale}, {Ford}, {Fulton},
  {Garza}, {Gomez}, {Graham}, {Greene}, {Haldeman}, {Hawkins}, {Haworth},
  {Haynes}, {Hidas}, {Hjelstrom}, {Howell}, {Hygelund}, {Lister}, {Lobdill},
  {Martinez}, {Mullins}, {Norbury}, {Parrent}, {Paulson}, {Petry}, {Pickles},
  {Posner}, {Rosing}, {Ross}, {Sand}, {Saunders}, {Shobbrook}, {Shporer},
  {Street}, {Thomas}, {Tsapras}, {Tufts}, {Valenti}, {Vander Horst}, {Walker},
  {White}, \& {Willis}}]{BBB2013}
{Brown}, T.~M., {et~al.} 2013, PASP, in press (arXiv: 1305.2437)

\bibitem[{{Cao} {et~al.}(2013){Cao}, {Gorbikov}, {Arcavi}, {Ofek}, {Gal-Yam},
  {Nugent}, \& {Kasliwal}}]{ATel5137}
{Cao}, Y., {Gorbikov}, E., {Arcavi}, I., {Ofek}, E., {Gal-Yam}, A., {Nugent},
  P., \& {Kasliwal}, M. 2013, The Astronomer's Telegram, 5137, 1

\bibitem[{{Cappa} {et~al.}(2004){Cappa}, {Goss}, \& {van der Hucht}}]{CGv2004}
{Cappa}, C., {Goss}, W.~M., \& {van der Hucht}, K.~A. 2004, \aj, 127, 2885

\bibitem[{{Cenko} {et~al.}(2006){Cenko}, {Fox}, {Moon}, {Harrison}, {Kulkarni},
  {Henning}, {Guzman}, {Bonati}, {Smith}, {Thicksten}, {Doyle}, {Petrie},
  {Gal-Yam}, {Soderberg}, {Anagnostou}, \& {Laity}}]{CFM2006}
{Cenko}, S.~B., {et~al.} 2006, \pasp, 118, 1396

\bibitem[{{Chevalier}(1998)}]{Chevalier1998}
{Chevalier}, R.~A. 1998, \apj, 499, 810

\bibitem[{{Chevalier} \& {Fransson}(2006)}]{CF2006}
{Chevalier}, R.~A., \& {Fransson}, C. 2006, \apj, 651, 381

\bibitem[{{Corsi} {et~al.}(2012){Corsi}, {Ofek}, {Gal-Yam}, {Frail},
  {Poznanski}, {Mazzali}, {Kulkarni}, {Kasliwal}, {Arcavi}, {Ben-Ami}, {Cenko},
  {Filippenko}, {Fox}, {Horesh}, {Howell}, {Kleiser}, {Nakar}, {Rabinak},
  {Sari}, {Silverman}, {Xu}, {Bloom}, {Law}, {Nugent}, \& {Quimby}}]{COG2012}
{Corsi}, A., {et~al.} 2012, \apjl, 747, L5

\bibitem[{{Crockett} {et~al.}(2007){Crockett}, {Smartt}, {Eldridge}, {Mattila},
  {Young}, {Pastorello}, {Maund}, {Benn}, \& {Skillen}}]{CSE2007}
{Crockett}, R.~M., {et~al.} 2007, \mnras, 381, 835

\bibitem[{{Dessart} {et~al.}(2012){Dessart}, {Hillier}, {Li}, \&
  {Woosley}}]{DHL2012}
{Dessart}, L., {Hillier}, D.~J., {Li}, C., \& {Woosley}, S. 2012, \mnras, 424,
  2139

\bibitem[{{Dolphin}(2000)}]{Dolphin2000}
{Dolphin}, A.~E. 2000, \pasp, 112, 1383

\bibitem[{{Eldridge} {et~al.}(2013){Eldridge}, {Fraser}, {Smartt}, {Maund}, \&
  {Crockett}}]{EFS2013}
{Eldridge}, J.~J., {Fraser}, M., {Smartt}, S.~J., {Maund}, J.~R., \&
  {Crockett}, R.~M. 2013, ArXiv: 1301.1975

\bibitem[{{Filippenko}(1997)}]{Filippenko1997}
{Filippenko}, A.~V. 1997, \araa, 35, 309

\bibitem[{{Filippenko} {et~al.}(1995){Filippenko}, {Barth}, {Matheson},
  {Armus}, {Brown}, {Espey}, {Fan}, {Goodrich}, {Ho}, {Junkkarinen}, {Koo},
  {Lehnert}, {Martel}, {Mazzarella}, {Miller}, {Smith}, {Tytler}, \&
  {Wirth}}]{FBM1995}
{Filippenko}, A.~V., {et~al.} 1995, \apjl, 450, L11

\bibitem[{{Hachinger} {et~al.}(2012){Hachinger}, {Mazzali}, {Taubenberger},
  {Hillebrandt}, {Nomoto}, \& {Sauer}}]{HMT2012}
{Hachinger}, S., {Mazzali}, P.~A., {Taubenberger}, S., {Hillebrandt}, W.,
  {Nomoto}, K., \& {Sauer}, D.~N. 2012, \mnras, 422, 70

\bibitem[{{Hill} {et~al.}(1998){Hill}, {Nicklas}, {MacQueen}, {Tejada}, {Cobos
  Duenas}, \& {Mitsch}}]{Hill98}
{Hill}, G.~J., {Nicklas}, H.~E., {MacQueen}, P.~J., {Tejada}, C., {Cobos
  Duenas}, F.~J., \& {Mitsch}, W. 1998, in Society of Photo-Optical
  Instrumentation Engineers (SPIE) Conference Series, Vol. 3355, Society of
  Photo-Optical Instrumentation Engineers (SPIE) Conference Series, ed.
  S.~{D'Odorico}, 375--386

\bibitem[{{Horesh} {et~al.}(2012){Horesh}, {Kulkarni}, {Fox}, {Carpenter},
  {Kasliwal}, {Ofek}, {Quimby}, {Gal-Yam}, {Cenko}, {de Bruyn}, {Kamble},
  {Wijers}, {van der Horst}, {Kouveliotou}, {Podsiadlowski}, {Sullivan},
  {Maguire}, {Howell}, {Nugent}, {Gehrels}, {Law}, {Poznanski}, \&
  {Shara}}]{HKF2012}
{Horesh}, A., {et~al.} 2012, \apj, 746, 21

\bibitem[{{Kamble} \& {Soderberg}(2013)}]{ATel5154}
{Kamble}, A., \& {Soderberg}, A. 2013, The Astronomer's Telegram, 5154, 1

\bibitem[{{Kasen}(2010)}]{Kasen2010}
{Kasen}, D. 2010, \apj, 708, 1025

\bibitem[{{Kasliwal} {et~al.}(2013){Kasliwal}, {Degenaar}, \&
  {Polishook}}]{ATel5151}
{Kasliwal}, M.~M., {Degenaar}, N., \& {Polishook}, D. 2013, The Astronomer's
  Telegram, 5151, 1
  
\bibitem[{{Kong} {et~al.}(2013){Kong}, {Li}, \& {Ofek}}]{ATel5210}
{Kong}, A.~K.~H., {Li}, K.~L., \& {Ofek}, E. 2013, The Astronomer's Telegram,
  5210, 1

\bibitem[{{Larkin} {et~al.}(2006){Larkin}, {Barczys}, {Krabbe}, {Adkins},
  {Aliado}, {Amico}, {Brims}, {Campbell}, {Canfield}, {Gasaway}, {Honey},
  {Iserlohe}, {Johnson}, {Kress}, {LaFreniere}, {Lyke}, {Magnone}, {Magnone},
  {McElwain}, {Moon}, {Quirrenbach}, {Skulason}, {Song}, {Spencer}, {Weiss}, \&
  {Wright}}]{LBK2006}
{Larkin}, J., {et~al.} 2006, in Society of Photo-Optical Instrumentation
  Engineers (SPIE) Conference Series, Vol. 6269, Society of Photo-Optical
  Instrumentation Engineers (SPIE) Conference Series

\bibitem[{{Law} {et~al.}(2009){Law}, {Kulkarni}, {Dekany}, {Ofek}, {Quimby},
  {Nugent}, {Surace}, {Grillmair}, {Bloom}, {Kasliwal}, {Bildsten}, {Brown},
  {Cenko}, {Ciardi}, {Croner}, {Djorgovski}, {van Eyken}, {Filippenko}, {Fox},
  {Gal-Yam}, {Hale}, {Hamam}, {Helou}, {Henning}, {Howell}, {Jacobsen},
  {Laher}, {Mattingly}, {McKenna}, {Pickles}, {Poznanski}, {Rahmer}, {Rau},
  {Rosing}, {Shara}, {Smith}, {Starr}, {Sullivan}, {Velur}, {Walters}, \&
  {Zolkower}}]{LKD2009}
{Law}, N.~M., {et~al.} 2009, \pasp, 121, 1395

\bibitem[{{Li} {et~al.}(2011){Li}, {Chornock}, {Leaman}, {Filippenko},
  {Poznanski}, {Wang}, {Ganeshalingam}, \& {Mannucci}}]{LCL2011}
{Li}, W., {Chornock}, R., {Leaman}, J., {Filippenko}, A.~V., {Poznanski}, D.,
  {Wang}, X., {Ganeshalingam}, M., \& {Mannucci}, F. 2011, \mnras, 412, 1473

\bibitem[{{Marion} {et~al.}(2013){Marion}, {Vinko}, {Kirshner}, {Foley},
  {Berlind}, {Bieryla}, {Bloom}, {Calkins}, {Challis}, {Chevalier}, {Chornock},
  {Culliton}, {Curtis}, {Esquerdo}, {Everett}, {Falco}, {France}, {Fransson},
  {Friedman}, {Garnavich}, {Leibundgut}, {Meyer}, {Smith}, {Soderberg},
  {Sollerman}, {Starr}, {Szklenar}, {Takats}, \& {Wheeler}}]{MVK2013}
{Marion}, G.~H., {et~al.} 2013, ArXiv: 1303.5482

\bibitem[{{Milisavljevic} {et~al.}(2013){Milisavljevic}, {Fesen}, {Pickering},
  {Miszalski}, {Buckley}, {Parrent}, {Marion}, {Silverman}, {Vinko}, {Wheeler},
  {Quimby}, {Jha}, {Mohamed}, {Kasliwal}, \& {Soderberg}}]{ATel5142}
{Milisavljevic}, D., {et~al.} 2013, The Astronomer's Telegram, 5142, 1

\bibitem[{{Nakar} \& {Sari}(2010)}]{NS2010}
{Nakar}, E., \& {Sari}, R. 2010, \apj, 725, 904

\bibitem[{{Ofek} {et~al.}(2012){Ofek}, {Laher}, {Law}, {Surace}, {Levitan},
  {Sesar}, {Horesh}, {Poznanski}, {van Eyken}, {Kulkarni}, {Nugent},
  {Zolkower}, {Walters}, {Sullivan}, {Ag{\"u}eros}, {Bildsten}, {Bloom},
  {Cenko}, {Gal-Yam}, {Grillmair}, {Helou}, {Kasliwal}, \& {Quimby}}]{oll+12}
{Ofek}, E.~O., {et~al.} 2012, \pasp, 124, 62

\bibitem[{{Piro} \& {Nakar}(2013)}]{PN2013}
{Piro}, A.~L., \& {Nakar}, E. 2013, \apj, 769, 67

\bibitem[{{Poznanski} {et~al.}(2012){Poznanski}, {Prochaska}, \&
  {Bloom}}]{PPB2012}
{Poznanski}, D., {Prochaska}, J.~X., \& {Bloom}, J.~S. 2012, \mnras, 426, 1465

\bibitem[{{Rabinak} \& {Waxman}(2011)}]{RW2011}
{Rabinak}, I., \& {Waxman}, E. 2011, \apj, 728, 63

\bibitem[{{Rayner} {et~al.}(2003){Rayner}, {Toomey}, {Onaka}, {Denault},
  {Stahlberger}, {Vacca}, {Cushing}, \& {Wang}}]{RTO2003}
{Rayner}, J.~T., {Toomey}, D.~W., {Onaka}, P.~M., {Denault}, A.~J.,
  {Stahlberger}, W.~E., {Vacca}, W.~D., {Cushing}, M.~C., \& {Wang}, S. 2003,
  \pasp, 115, 362

\bibitem[{{Richmond} {et~al.}(1996){Richmond}, {van Dyk}, {Ho}, {Peng}, {Paik},
  {Treffers}, {Filippenko}, {Bustamante-Donas}, {Moeller}, {Pawellek},
  {Tartara}, \& {Spence}}]{RvH1996}
{Richmond}, M.~W., {et~al.} 1996, \aj, 111, 327

\bibitem[{{Sault} {et~al.}(1995){Sault}, {Teuben}, \& {Wright}}]{STW1995}
{Sault}, R.~J., {Teuben}, P.~J., \& {Wright}, M.~C.~H. 1995, in Astronomical
  Society of the Pacific Conference Series, Vol.~77, Astronomical Data Analysis
  Software and Systems IV, ed. R.~A. {Shaw}, H.~E. {Payne}, \& J.~J.~E.
  {Hayes}, 433

\bibitem[{{Schlafly} \& {Finkbeiner}(2011)}]{SF2011}
{Schlafly}, E.~F., \& {Finkbeiner}, D.~P. 2011, \apj, 737, 103

\bibitem[{{Simcoe} {et~al.}(2013){Simcoe}, {Burgasser}, {Schechter}, {Fishner},
  {Bernstein}, {Bigelow}, {Pipher}, {Forrest}, {McMurtry}, {Smith}, \&
  {Bochanski}}]{SBS2013}
{Simcoe}, R.~A., {et~al.} 2013, \pasp, 125, 270

\bibitem[{{Soderberg} {et~al.}(2008){Soderberg}, {Berger}, {Page}, {Schady},
  {Parrent}, {Pooley}, {Wang}, {Ofek}, {Cucchiara}, {Rau}, {Waxman}, {Simon},
  {Bock}, {Milne}, {Page}, {Barentine}, {Barthelmy}, {Beardmore}, {Bietenholz},
  {Brown}, {Burrows}, {Burrows}, {Byrngelson}, {Cenko}, {Chandra}, {Cummings},
  {Fox}, {Gal-Yam}, {Gehrels}, {Immler}, {Kasliwal}, {Kong}, {Krimm},
  {Kulkarni}, {Maccarone}, {M{\'e}sz{\'a}ros}, {Nakar}, {O'Brien}, {Overzier},
  {de Pasquale}, {Racusin}, {Rea}, \& {York}}]{SBP2008}
{Soderberg}, A.~M., {et~al.} 2008, \nat, 453, 469

\bibitem[{{Stritzinger} {et~al.}(2009){Stritzinger}, {Mazzali}, {Phillips},
  {Immler}, {Soderberg}, {Sollerman}, {Boldt}, {Braithwaite}, {Brown}, {Burns},
  {Contreras}, {Covarrubias}, {Folatelli}, {Freedman}, {Gonz{\'a}lez}, {Hamuy},
  {Krzeminski}, {Madore}, {Milne}, {Morrell}, {Persson}, {Roth}, {Smith}, \&
  {Suntzeff}}]{SMP2009}
{Stritzinger}, M., {et~al.} 2009, \apj, 696, 713

\bibitem[{{Taubenberger} {et~al.}(2006){Taubenberger}, {Pastorello}, {Mazzali},
  {Valenti}, {Pignata}, {Sauer}, {Arbey}, {B{\"a}rnbantner}, {Benetti}, {Della
  Valle}, {Deng}, {Elias-Rosa}, {Filippenko}, {Foley}, {Goobar}, {Kotak}, {Li},
  {Meikle}, {Mendez}, {Patat}, {Pian}, {Ries}, {Ruiz-Lapuente}, {Salvo},
  {Stanishev}, {Turatto}, \& {Hillebrandt}}]{TPM2006}
{Taubenberger}, S., {et~al.} 2006, \mnras, 371, 1459

\bibitem[{{Tully} {et~al.}(2009){Tully}, {Rizzi}, {Shaya}, {Courtois},
  {Makarov}, \& {Jacobs}}]{TRS2009}
{Tully}, R.~B., {Rizzi}, L., {Shaya}, E.~J., {Courtois}, H.~M., {Makarov},
  D.~I., \& {Jacobs}, B.~A. 2009, \aj, 138, 323

\bibitem[{{Tuthill} {et~al.}(1999){Tuthill}, {Monnier}, \& {Danchi}}]{TMD1999}
{Tuthill}, P.~G., {Monnier}, J.~D., \& {Danchi}, W.~C. 1999, \nat, 398, 487

\bibitem[{{Valenti} {et~al.}(2008){Valenti}, {Elias-Rosa}, {Taubenberger},
  {Stanishev}, {Agnoletto}, {Sauer}, {Cappellaro}, {Pastorello}, {Benetti},
  {Riffeser}, {Hopp}, {Navasardyan}, {Tsvetkov}, {Lorenzi}, {Patat}, {Turatto},
  {Barbon}, {Ciroi}, {Di Mille}, {Frandsen}, {Fynbo}, {Laursen}, \&
  {Mazzali}}]{VET2008}
{Valenti}, S., {et~al.} 2008, \apjl, 673, L155

\bibitem[{{Valenti} {et~al.}(2011){Valenti}, {Fraser}, {Benetti}, {Pignata},
  {Sollerman}, {Inserra}, {Cappellaro}, {Pastorello}, {Smartt}, {Ergon},
  {Botticella}, {Brimacombe}, {Bufano}, {Crockett}, {Eder}, {Fugazza},
  {Haislip}, {Hamuy}, {Harutyunyan}, {Ivarsen}, {Kankare}, {Kotak}, {Lacluyze},
  {Magill}, {Mattila}, {Maza}, {Mazzali}, {Reichart}, {Taubenberger},
  {Turatto}, \& {Zampieri}}]{VFB2011}
---. 2011, \mnras, 416, 3138

\bibitem[{{Vogt} {et~al.}(1994){Vogt}, {Allen}, {Bigelow}, {Bresee}, {Brown},
  {Cantrall}, {Conrad}, {Couture}, {Delaney}, {Epps}, {Hilyard}, {Hilyard},
  {Horn}, {Jern}, {Kanto}, {Keane}, {Kibrick}, {Lewis}, {Osborne},
  {Pardeilhan}, {Pfister}, {Ricketts}, {Robinson}, {Stover}, {Tucker}, {Ward},
  \& {Wei}}]{VAB1994}
{Vogt}, S.~S., {et~al.} 1994, in Society of Photo-Optical Instrumentation
  Engineers (SPIE) Conference Series, Vol. 2198, Society of Photo-Optical
  Instrumentation Engineers (SPIE) Conference Series, ed. D.~L. {Crawford} \&
  E.~R. {Craine}, 362

\bibitem[{{Wizinowich} {et~al.}(2006){Wizinowich}, {Chin}, {Johansson},
  {Kellner}, {Lafon}, {Le Mignant}, {Neyman}, {Stomski}, {Summers}, {Sumner},
  \& {van Dam}}]{WCJ2006}
{Wizinowich}, P.~L., {et~al.} 2006, in Society of Photo-Optical Instrumentation
  Engineers (SPIE) Conference Series, Vol. 6272, Society of Photo-Optical
  Instrumentation Engineers (SPIE) Conference Series

\bibitem[{{Yaron} \& {Gal-Yam}(2012)}]{YG2012}
{Yaron}, O., \& {Gal-Yam}, A. 2012, \pasp, 124, 668

\bibitem[{{Yoon} {et~al.}(2012){Yoon}, {Gr{\"a}fener}, {Vink}, {Kozyreva}, \&
  {Izzard}}]{YGV2012}
{Yoon}, S.-C., {Gr{\"a}fener}, G., {Vink}, J.~S., {Kozyreva}, A., \& {Izzard},
  R.~G. 2012, \aap, 544, L11

\end{thebibliography}
\end{document}